# Bulk Ferroelectric Heterostructures for High Temperature Lead-Free Piezoelectrics

Yizhe Li[1,2,8]*, Ziqi Yang[1,2,8], Ying Chen[1,2], Zhenbo Zhang[3], Yun-Long Tang[4], Matthew Smith[1], Matthew Lindley[1,2], Xuezhen Cao[1,2], David G. Hopkinson[5], Andrew J. Bell[6], Steven J. Milne[6], Antonio Feteira[7], Sarah J. Haigh[1,2], Alexander S. Eggeman[1], Juncheng Pan[1,2], Jiajun Shi[1,2], David A. Hall[1,2]*

[1]Department of Materials, University of Manchester, M13 9PL, Manchester, UK. [2]Henry Royce Institute, University of Manchester, M13 9PL, Manchester, UK. [3]Center for Adaptive System Engineering, School of Creativity and Arts, ShanghaiTech University, Shanghai, 201210, China. [4]Shenyang National Laboratory for Materials Science, Institute of Metal Research, Chinese Academy of Sciences, Wenhua Road 72, Shenyang 110016, China [5]electron Physical Science Imaging Centre, Diamond Light Source, Ltd., Harwell Science & Innovation Campus, Didcot, OX11 0DE,UK [6]School of Chemical and Process Engineering, University of Leeds, LS2 9JT, Leeds, UK. [7]Materials and Engineering Research Institute, Sheffield Hallam University, Sheffield, S1 1WB, UK. [8]These authors contributed equally: Yizhe Li and Ziqi Yang. *e-mail: yizhe.li@manchester.ac.uk; david.a.hall@manchester.ac.uk

## Abstract

The remarkable exploitation of valence and lattice mismatch in epitaxial ferroelectric heterostructures generates physical effects not classically expected for perovskite oxides, such as the 2D electron gas and polar skyrmions [1]. However the widespread application of these interfacial properties and functionalities is impeded by the ultrathin layered structure and essential presence of underlying lattice-matched substrates for the deposition of epitaxial thin films. Here, we report a bottom-up pathway to synthesize bulk ferroelectric heterostructures (BFH) with periodic composition fluctuation (8 nm in wavelength) using elemental partitioning by cation diffusion, providing opportunities to exploit the novel characteristics of hetero-epitaxial oxide thin films in bulk materials. The exemplar monolithic $BiFeO_3$-$BaTiO_3$ BFH ceramics described herein share common features with their thin film heterostructure counterparts, which facilitates the control and stabilisation of ferroelectric polarisation along with a significant enhancement in Curie temperature, $T_C$, and functional performance[2–4]. The developed BFH ceramics exhibit a record $T_c$ (up to 824 °C) and a high piezoelectric coefficient ($d_{33}$ = 115 pC N$^{-1}$), in comparison with other perovskite or non-perovskite solid solutions, providing practical and sustainable solutions for emergent high temperature (> 400 °C) piezoelectric sensing, actuation and energy conversion applications[5]. By creating BFH ceramics using different electromechanical boundary conditions, distinct morphologies of aliovalent A-site cation segregated regions along with different types of ferroelectric order are achieved, governed by the local electrostatic potential. This straightforward formation mechanism provides unprecedented control over the local ferroelectric ordering and domain stabilisation in BFH ceramics; it also paves the way to explore new types of functionality, beyond those achievable in both bulk ferroelectrics and thin film heterostructures.

Since the discovery of piezoelectricity in 1880, such materials have been employed extensively for applications in electromechanical sensors, actuators and energy conversion [6,7]. The dominant lead zirconate titanate (PZT) ceramics and high performance relaxor-$PbTiO_3$ single crystals exhibit incomparable sensitivity, accuracy, efficiency and operational simplicity in comparison with competing technologies. However, the development of piezoelectric materials with high sensitivity and efficiency for high temperature applications has remained as a significant challenge since the 1970s [8,9]. On the other hand, there are growing demands for on-line process monitoring, continuous structural integrity observation and actuation applications in the harsh and high temperature environments associated with the energy generation, petrol refinery, mining, metallurgy, automotive, aviation, and aerospace sectors[10]. As an exemplar application, it was noted that within the EU for 2010, an economic saving of € 2.5 billion per annum could potentially be achieved by a 10% reduction in the forced outage time of electricity power plants, if on-line ultrasonic sensing systems operating at 580 °C could be deployed for continuous condition monitoring of superheated steam line welds (supercritical steam at 538 to 565 °C) [11,12].

It is worth noting that the maximum operation temperatures for ferroelectric type piezoelectrics are determined by their depolarisation temperatures, $T_d$, above which the randomization of the pre-defined domain configuration produced by 'poling' occurs, leading to a rapid loss of remanent polarization and piezoelectricity even before reaching the Curie temperature, $T_c$. The difference between $T_d$ and $T_c$ is typically of the order of 100 °C[10]. The highest $T_c$ values for commercial PZT ceramics are typically in the region of 350 °C, with a corresponding operational temperature limit of 250 °C being imposed to avoid the rapid degradation in piezoelectric performance on approaching $T_c$ (Fig. 1 a,b). These are far from the benchmark temperature of 580 °C, which would potentially enhance the capability of piezoceramics to encompass a much broader range of high temperature applications. Furthermore, lead-based piezoelectric ceramics such as PZT are also at risk due to environmental concerns and the progressive development of legislation for the substitution or removal of lead (Pb) in electronic devices[13].

Research on new or improved high temperature piezoelectric materials has focused mainly on the search for novel compositions and chemical modification of existing systems. In this regard, all of the previously identified high temperature polycrystalline piezoelectrics exhibit disadvantages, which impede their potential for widespread applications or industry scale-up. For example, piezoceramics based on tungsten bronze-structured lead metaniobate ($Pb_2Nb_2O_6$) and perovskite-structured $BiScO_3$-$PbTiO_3$ (BS-PT) exhibit useful piezoelectric performance with maximum operation temperatures up to 300-350 °C (Fig. 1 a,b)[14,15], but concerns over the use of lead and the prohibitive cost of $Sc_2O_3$ have damaged their prospects for further commercialization. As another high temperature candidate material, bismuth layer-structured ferroelectrics (BLSF, $T_c$ > 400°C) tend to exhibit low piezoelectric coefficients, typically around 20 pC N$^{-1}$ compared with 200 to 500 pC N$^{-1}$ for PZT (Fig. 1 a,b)[16,17]. This suggests that such materials are generally suitable only for passive applications in piezoelectric sensors and require additional impedance matching circuits and signal amplification to compensate for their low sensitivity[10]. Further development of $BiFeO_3$-$BaTiO_3$ (BF-BT) and similar perovskite solid solutions yields enhanced $T_c$ up to about 500 °C with comparable piezoelectric properties to those of hard PZTs[18]. Their maximum operation temperatures for continuous deployment are still below 400 °C, due to the rapid depolarisation above $T_d$ (Fig. 1e). Additional doping or chemical modification of the existing compositions normally enhance the room temperature piezoelectric performance by sacrificing the magnitude of $T_c$, i.e. decrease in the upper limit of the application temperature. Alternatively, the ferroelectricity of epitaxial heterostructures (few to tens of nm) can be enhanced and stabilised with the increment in $T_c$ by means of biaxial strain (> 1%) engineering[19], which is not transferable to bulk oxide materials with the absence of such high strain tolerance and inevitable strain relaxation. The ultrathin thickness of epitaxial thin films deposited on rigid substrates also restricts the output power level, resonance frequency and integration of the associated devices for active piezoelectric applications. In order to meet the urgent demands for high temperature piezoelectric applications, a unique method to design new types of high temperature ferroelectrics and piezoelectrics taking advantage of both epitaxial heterostructures and bulk materials would be a desirable pathway to unlock the full potential of ferroelectrics.

In the present study, we have established a novel approach to stabilise the ferroelectricity and piezoelectricity with enhanced $T_c$, via a controllable bottom-up synthesis for bulk ferroelectric heterostructures (BFH) to incorporate the exploitation of valence and lattice mismatch, similar to those of epitaxial thin film counterparts. The proposed synthesis route harnesses the immiscible nature of related oxide solid solutions and exploits the local elemental partitioning as a mechanism for the directional regulation of polarisation continuity and aliovalent elemental segregation by means of simple thermal ageing and domain engineering. The developed BF-BT BFH ceramics yield a $T_c$ up to 824 °C (Fig. 1c), room temperature piezoelectric coefficient, $d_{33}$=115 pC N$^{-1}$, and thickness coupling factor, $k_p$=0.37 (Fig. 1a,b), which exhibit superior sensitivity and efficiency comparing with other high temperature piezoelectric candidates capable of operating at 580 °C. The high temperature piezoelectric performances of BF-BT BFH ceramics ($d_{33}$ > 400 pC N$^{-1}$ and $k_{33}$ > 0.4) are comparable with those of commercial PZT ceramics (Fig. 1d). Short-term operation of BF-BT BFH ceramics up to 800 °C is demonstrated by the ex-situ piezoelectric coefficient ($d_{33}$) measurements after exposing the samples to specified temperatures (25-850 °C) for 30 min (details in Methods). The BF-BT BFH ceramics also exhibit exceptional long term stability at the benchmark temperature of 580 °C with less than 3% loss of the piezoelectric charge coefficient ($d_{33}$) after 1000 hours (Fig. 1f). The successful exploitation of the same approach to other solid solutions of $BiFeO_3$-$SrTiO_3$ and $Na_{0.5}Bi_{0.5}TiO_3$-$NaNbO_3$ (Extended Data Fig. 1) also serves to illustrate the potential of this novel approach for materials engineering, both in $BiFeO_3$-based systems and in a wide range of other related electroceramic solid solutions that may exhibit tendencies for immiscibility.

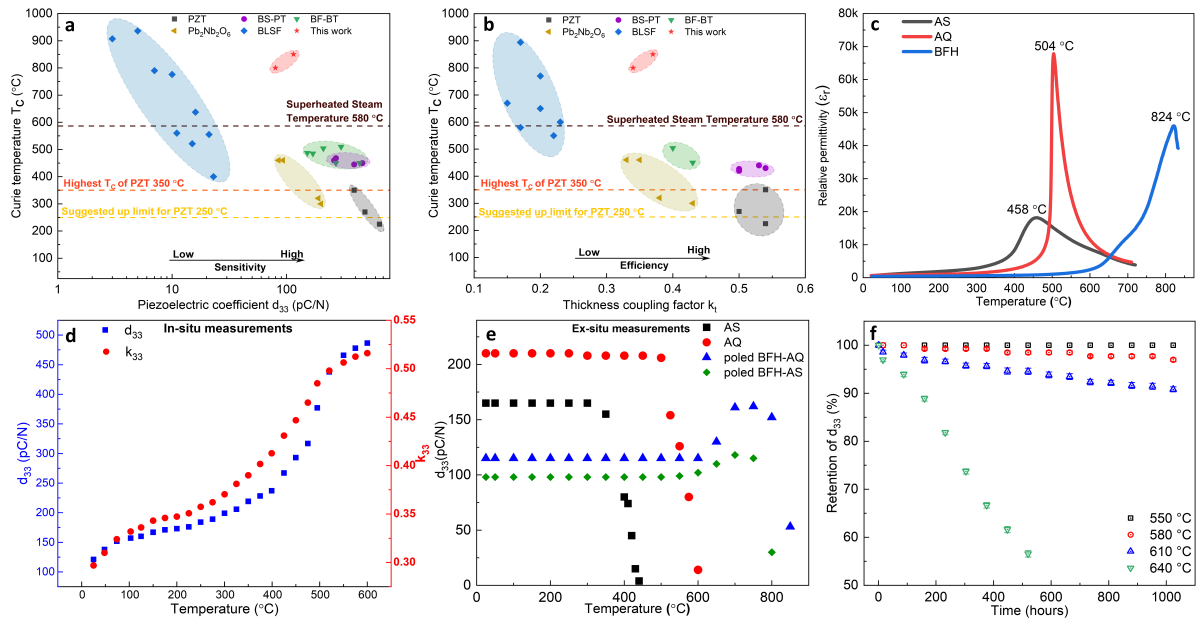

Fig. 1 **Electrical properties and associated stability of 0.7BF-0.3BT BFH ceramics.** Comparison of **a**, piezoelectric coefficient ($d_{33}$) and **b,** thickness coupling factor ($k_t$) between the BFH ceramics and other perovskite and non-perovskite ceramics, including PZT, BS-PT, BF-BT, $Pb_2Nb_2O_6$ and BLSF. **c,** temperature dependence of relative permittivity. **d**, temperature dependence of $d_{33}$ and coupling factor $k_{33}$ determined by resonance method. **e,** ex-situ measurement of $d_{33}$ after dwelling at various temperatures for as sintered (AS), air quenched (AQ) and poled BFH samples. **f,** retention of $d_{33}$ after dwelling at various temperatures for poled BFH samples with different duration up to 1000 hours.

Samples of (1-x)BF-xBT with various compositions (x=0.25, 0.3, 0.35, 0.4) were initially homogenised using planetary ball milling and synthesised by solid state reaction, followed by conventional high temperature sintering (further details given in Methods section). Subsequent heat treatment processes, involving air quenching (AQ) and thermal ageing (TA) (Fig 2. a), were used to homogenise the solid solution, and activate and control nanoscale elemental partitioning respectively. It is worth mentioning that the AQ solution treatment process, which serves to enhance the normal ferroelectric (FE) state and exclude the effects of uncontrolled elemental partitioning on microstructure and ferroelectric ordering during slow cooling after sintering, can be considered as optional. In this respect, solely thermal ageing of as-sintered (AS), slow-cooled BF-BT ceramics can also be used to create BFH ceramics with enhanced $T_c$ but somewhat impaired functional performance. However, the crack-free microstructure and enhanced fracture toughness of AQ samples indicates exceptional thermal shock resistance of BF-BT ceramics (Extended Data Fig. 2), indicating that the AQ process is a feasible and preferable method to manufacture BF-BT ceramics having enhanced piezoelectric performance[20]. The thermal stresses induced by quenching could be further mitigated by reducing the magnitude of the temperature change during quenching (e.g. from 900 °C to 400 °C) and tempering during thermal ageing.

Taking the 0.7BF-0.3BT composition as an exemplar pseudo-binary solid solution, transmission electron microscopy (TEM) analysis of the ceramic at different stages of the heat treatment process shows profound changes in the microstructure for both unpoled AS and poled AQ samples (Fig. 2b,d) after thermal ageing (Fig. 2c,e). Thermal ageing of the AS samples for 200 h at 500 °C (above the Curie temperature of the AS sample, 458 °C, Fig. 2a), induced the formation of a relatively disordered domain configuration with underlying nanoscale labyrinthine features (Fig 2.c and Extended Data Fig. 3), identified here as *bulk ferroelectric heterostructures* (BFH) with random local field. The corresponding ferroelectric polarisation-electric field (P-E) hysteresis loop of the AS BF-BT ceramic evolved into a slim loop for BFH-AS, exhibiting antiferroelectric-like characteristics, after thermal ageing (Fig. 2f). In contrast, the broad domain structure obtained after poling of the AQ (normal FE) sample (Fig. 2d) was retained in the poled BFH-AQ ceramic after thermal ageing below $T_c$ (Fig. 2a), but their integrity and continuity were disturbed by locally developed wave-like nanoscale BFH patterns induced by the ageing process (Fig. 2e and Extended Data Fig. 4). The corresponding asymmetric P-E loop of the poled BFH-AQ ceramic is shifted significantly along the electric field axis with an outstanding internal bias field ($E_{ib}$=7.5 MV m$^{-1}$), in comparison with the saturated but symmetric P-E loop of the AQ (normal FE) samples. From this work, three different types of ferroelectric characteristics and distinct microstructures were achieved by controllable bottom-up synthesis of BFH, by modifying the heat treatment parameters of a single BF-BT solid solution and without the need for conventional compositional modification or strain engineering.

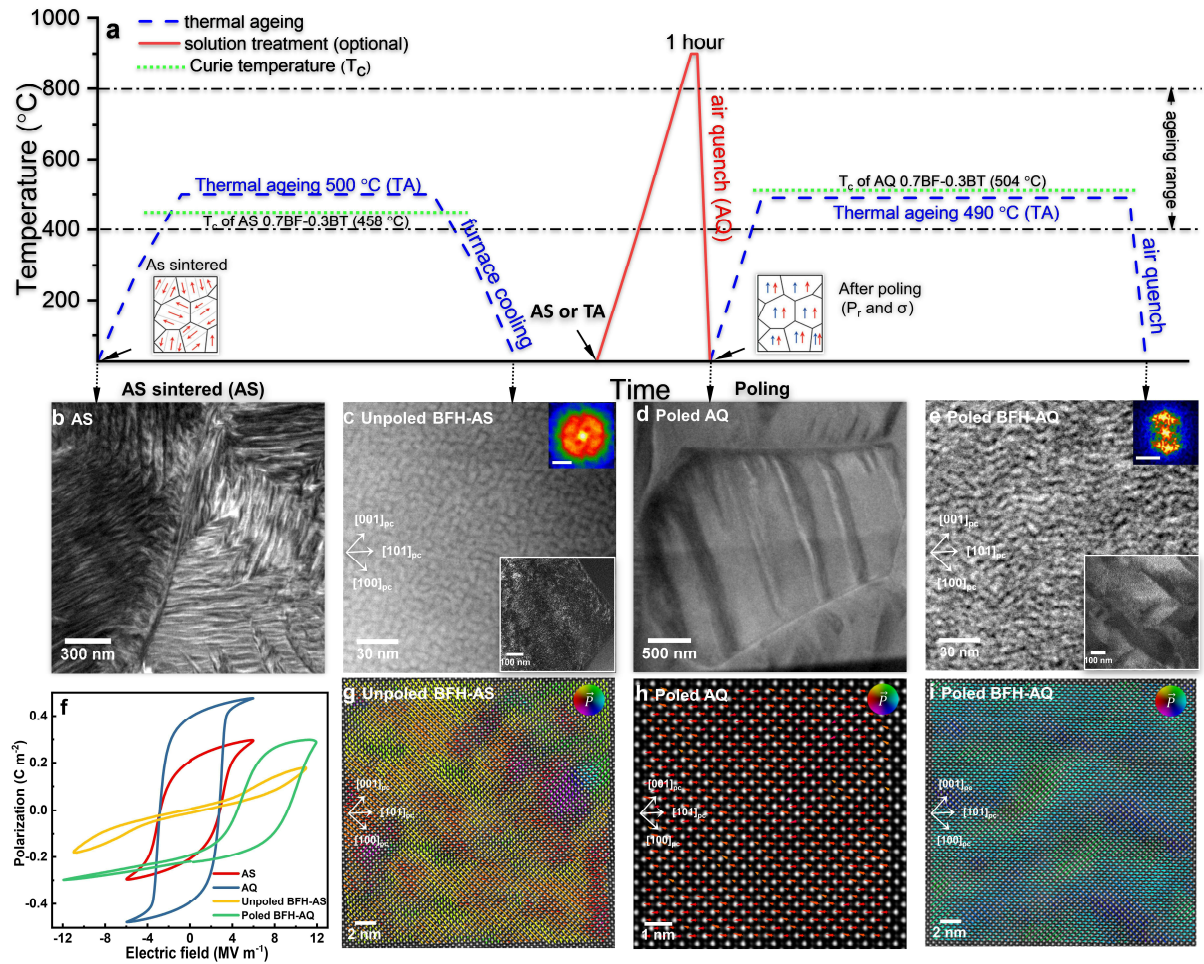

Fig. 2 **Heat treatment procedures and the corresponding evolution of microstructure and ferroelectric behaviour for 0.7BF-0.3BT BFH ceramics. a**, Temperature profile of the heat treatment consisting of thermal ageing (400 °C- 800 °C) and solution treatment followed by air quenching (1 hour dwell at 900 °C followed by rapid cooling to room temperature). Additional poling by an applied electric field can be used prior to thermal ageing to introduce a remnant polarisation ($P_r$) state and associated residual stress (σ). **b and d**, Bright field TEM images of AS and poled AQ ceramics (normal ferroelectrics) showing lamellar domain structures. **c**, HAADF image of BFH-AS ceramics viewed along [010]$_{pc}$ zone axis revealing the development of modulated nanoscale features with atomic number (Z) contrast variation, the top inset shows the fast Fourier transform (FFT) of **c**, and the bottom inset shows dark field TEM image of local polar nano regions. **e**, HAADF image of poled BFH-AQ ceramics viewed along [010]$_{pc}$ zone axis revealing the development of orientated wave-like patterns within a single domain, the top inset shows the fast Fourier transform (FFT) of **e**, and the bottom inset shows retention of broad domain structure after thermal ageing in poled BFH-AQ sample. **f**, P-E loops of AS, AQ, BFH-AS and poled BFH-AQ BF-BT ceramics corresponding to the microstructures shown in **b-e**. **g-i,** atomic-resolution HAADF images with the overlaid atomic polarisation mappings of BFH-AS, AQ and poled BFH-AQ samples.

High-angle annular dark-field (HAADF) – STEM, energy dispersive X-ray spectroscopy (EDS) and electron energy loss spectroscopy (EELS) imaging studies were conducted to further investigate the nano-scale patterns observed in TEM images of the BFH-AS and BFH-AQ samples. The collection of imaging and spectroscopic evidences reveal that these nano-scale patterns relate to local elemental partitioning regions with Ba$^{2+}$ and Bi$^{3+}$ segregation as shown in the HAADF STEM Z-contrast images (Fig. 2c,e). The elemental partitioning exists despite a coherent perovskite lattice structure extending across the whole region (Extended Data Fig. 3d and Extended Data Fig. 4d). In comparison with the long range ordering of ferroelectricity in the AQ sample with rhombohedral symmetry (Fig. 2h), the presence of aliovalent elemental partitioning of $Ba^{2+}$ and $Bi^{3+}$ significantly disturbs the ferroelectric ordering and related local electric field and polarisation continuity in both BFH-AS and poled BFH-AQ samples (Fig. 2g,i). The results obtained by atomic scale polarisation mapping of the BFH-AS samples reveal the presence of random local field orientations with continuous polarisation rotation (Fig. 2g and Extended Data Fig. 5). In contrast, an alternating convergent and divergent polarisation distribution centred along the $[111]_{pc}$ (pseudo-cubic notation) direction was observed in poled BFH-AQ samples (Fig. 2i and Extended Data Fig. 6), which is consistent with the modulated spatial arrangement of nanoscale patterns as shown in Fig. 2e. The relationship between the polarisation orientation distribution and elemental partitioning will be discussed later with further detailed elemental analysis.

The formation of nanoscale patterns with modulated spatial arrangements in both BFH-AS and poled BFH-AQ samples (Fig. 2c,e) are governed by crystallographic elastic anisotropy (particularly between polar and non-polar directions) and the requirement for minimization of the coherent lattice strain energy[22]. The presence of wave-like Bi- and Ba-enriched regions with 2-fold symmetry was observed in both the as sintered BFH-AS and poled BFH-AQ samples (Fig. 2c,e ) when viewed along zone axis $[010]_{pc}$. In the BFH-AS samples, the orientated growth of the wavelike patterns with 2 fold symmetry and cuboidal features show a general tendency for alignment along both $[100]_{pc}$ and $[001]_{pc}$ directions, indicated by the diffuse 4-fold side lobes surrounding the central diffraction spot in the fast Fourier transform (FFT) of the HAADF image (top inset of Fig. 2c)[23], which is attributed to the absence of the long range ordered domains at the thermal ageing temperature of 500 °C, which is greater than the ferroelectric Curie temperature ($T_c$ = 458 °C) of the AS samples. In contrast, the morphological alignment of the elementally partitioned, phase separated regions in the poled BFH-AQ sample is mainly along $[111]_{pc}$ or $[\bar{1}11]_{pc}$ directions (corresponding to $[101]_{pc}$ and $[\bar{1}01]_{pc}$ within the plane view) in different parent domains (Fig. 2i and Fig. 3b,e). The enhanced aspect ratio and exclusive <111> growth orientation of the nanoscale pattern within an individual parent domain (Fig 2. c) yields 2-fold side lobes in the FFT (inset Fig 2. c), which is attributed to the preferred polarisation direction and associated elastic strain induced by pre-poling. Furthermore, a characteristic diffuse spinodal ring[25] was also evident in the selected area electron diffraction pattern (SAED, inset of Extended Data Fig 4. a). The wavelength of the periodic compositional fluctuation or the size of the nano features developed in the present study is around 8 nm according to the evaluation of the spinodal ring in SAED and the positions of the side lobes in the HAADF FFT (Extended Data Fig. 7).

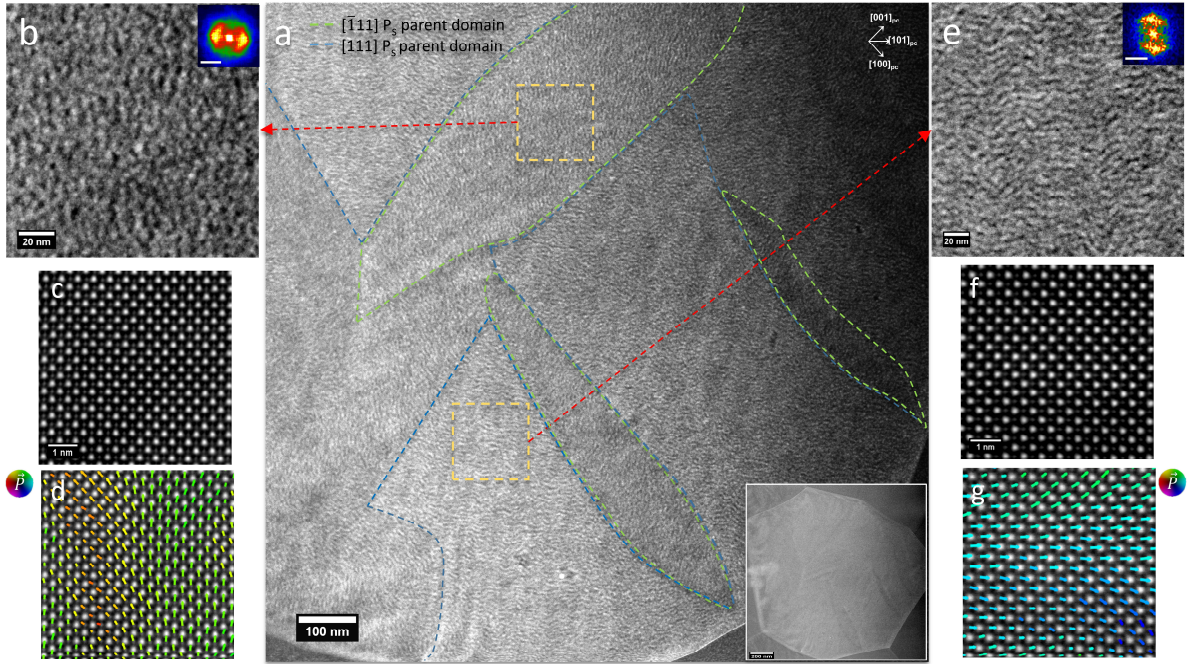

Fig. 3 **Spatial arrangements of the wave-like features in 0.7BF-0.3BT poled BFH-AQ ceramic according to the local spontaneous polarisation direction. a,** HAADF image of [010] orientated grain contains two types of parent domain variants with polarisation along $[111]_{pc}$ and $[\bar{1}11]_{pc}$ directions (corresponding to $[101]_{pc}$ and $[\bar{1}01]_{pc}$ directions within the plane view respectively); the inset in **a** shows the full view of the investigated grain. **b,** HAADF image for the identical region (dashed square) marked in the parent domain with polarisation along $[\bar{1}11]_{pc}$ in **a**, showing $[\bar{1}11]_{pc}$ preferred growth of the wave-like features with associated 2-fold side lobes along the $[111]_{pc}$ direction in the FFT inset. **c,d** atomic-resolution HAADF image and associated polarisation vector mapping for the local region in **b** confirms the divergent polarisation distribution centred along $[\bar{1}11]_{pc}$. **e,** HAADF image for the identical region (dashed square) marked in the parent domain with polarisation along $[111]_{pc}$ in **a**, showing $[101]_{pc}$ preferred growth of the wave-like features with associated 2-fold side lobes along $[\bar{1}11]_{pc}$ in the FFT inset. **f,g** atomic-resolution HAADF image and associated polarisation vector mapping for the local region in **e** confirms the divergent polarisation distribution centred along $[111]_{pc}$. The scale bars are 0.2 nm$^{-1}$ for both insets in **b,e**.

Further HAADF tomographic 3D reconstruction of the local region in poled BFH-AQ samples reveals the percolation between Ba$^{2+}$ and Bi$^{3+}$ enriched regions by forming cross-linked lamellar networks (Fig. 4a), in comparison with the 2D wave-like features view along the $[010]_{pc}$ zone axis (Fig. 3a). These results provide the microstructural foundation, analogous to the ferroelectric epitaxial heterostructures or superlattices in a bulk polycrystalline ceramic form, for the unique macroscopic properties of BFH. At the interfaces between these elementally segregated structures, the alternating divergent and convergent orientation of projected polarisation vectors is modulated along with the alternating enrichment of Ba$^{2+}$ and Bi$^{3+}$, according to the atomic EDS mapping (Fig. 4e). The standardless EDS mapping results (Extended Data Fig. 8) indicate that the amplitude of the A-site concentration variations for Ba$^{2+}$ and Bi$^{3+}$ between the element separated regions is approximately 20 at% (4 at% in total of the ABO$_3$ formula unit). The STEM-EDS data reveals no visible segregation of Fe, Ti and O to either separate region (Extended Data Fig. 8) although the overlap of the characteristic X-ray peaks of Ba (Lα$_1$ 4.465 keV) and Ti (Kα$_1$ 4.508 keV)[26] may impede the interpretation of EDS results with respect to segregation of the B-site cations.

Complementary mapping by EELS (Extended Data Fig. 9) further illustrates the elemental distributions of Ba, Ti, Fe and O. EELS struggled to detect Bi due to the high energy loss and delayed edge structure but provides superior sensitivity for light elements. Analysis of sum spectra for higher and lower intensity regions of the HAADF image confirms the STEM-EDS result of Ba segregation and also shows the Ba enrichment (Bi depletion) is associated with a slight Ti enrichment and Fe depletion (Extended data Fig. 9a-e). The disproportionate B-site element segregation only partially compensates the charge imbalance induced by respective aliovalent Ba$^{2+}$ and $Bi^{3+}$ exchange with the formation of associated $Bi^{\bullet}_{Ba}$ and $Ba'_{Bi}$ point defects in adjacent regions showing segregation. The possible variance of oxidation states for Fe and Ti cations as a charge balancing mechanism is also ruled out by the almost identical chemical shifts and the edge onset EELS spectra for O, Fe and Ti (Extended data Fig. 9f-h) for both Ba-enriched and Bi-enriched regions[27]. Further complete charge compensation is achieved by the development of alternating divergent and convergent polarisation distribution (i.e. charged head-to-head (H-H) and tail-to-tail (T-T) domain wall structures) during thermal ageing, which will be further discussed in the following section.

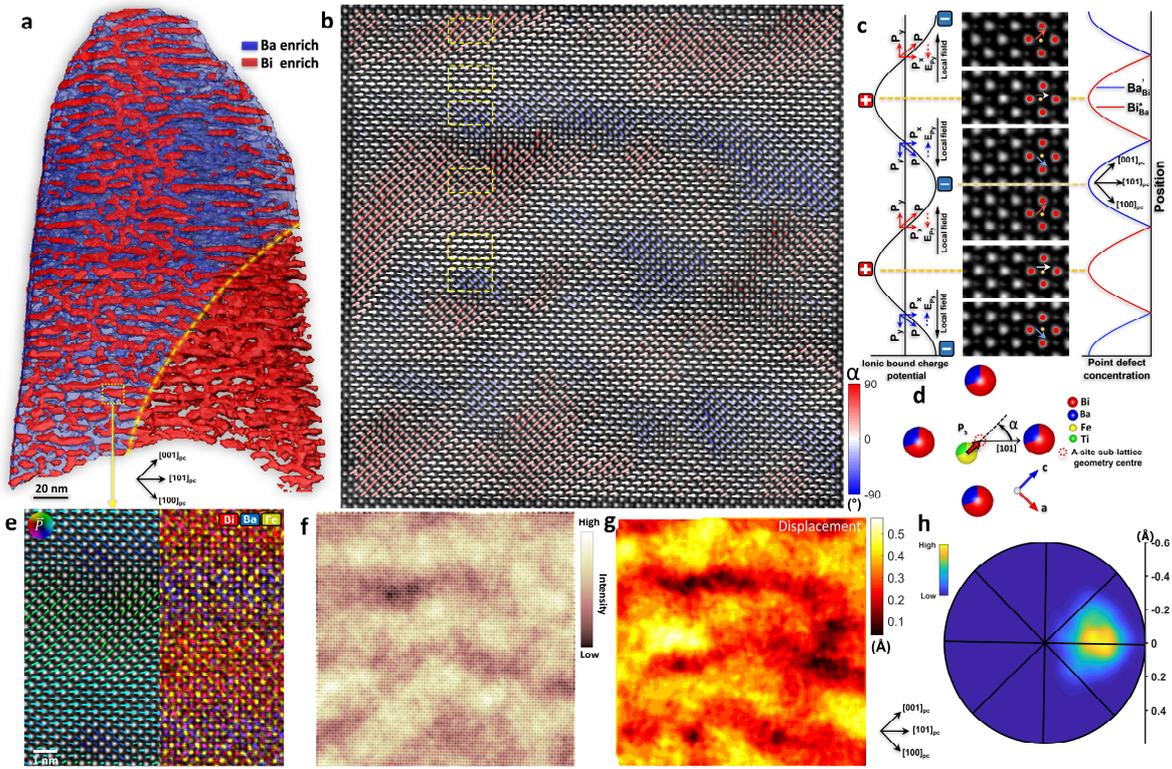

Fig. 4. **HAADF-STEM tomography and correlation between polarisation and element distributions in poled BFH-AQ BF-BT ceramics**. **a**, HAADF tomographic 3D reconstruction of the local region form the poled BFH(AQ) sample shows cross-linked lamellar networks of Bi- and Ba- enriched phases percolation. **b**, Atomic scale polarisation vector mapping with colour indexing local polarisation vectors by the angle, α, relative to the $[101]_{pc}$ direction. **c**, Schematic illustration of the variations in local point defect concentration and the corresponding local electric field potential induced by space charge due to preferential segregation of $Bi^{3+}$ and $Ba^{2+}$ cations. The atomic resolution HAADF-STEM images sampled within the yellow-dashed rectangle regions in **b** illustrate the periodic reorientation of polarisation due to minimisation of interfacial elastic and electrostatic energies. **d**, Schematic illustration of the misorientation angle, α, between the local polarisation vector and the $[101]_{pc}$ direction. **e**, the local alternating convergent and divergent polarisation distribution mirrored with the associated atomic-resolution EDX mapping showing modulated $Bi^{3+}$ and $Ba^{2+}$ partitioning. **f**, Intensity distribution, **g**, mapping of the B-site cation displacement and **h**, polar plot of polarisation vector orientation distribution for the same region shown in **b**.

By colour indexing the local polarisation vectors according to the angle, α, relative to the $[101]_{pc}$ direction, HAADF image intensity and mapping the B-site cation displacement magnitude (as illustrated in Fig 4. c,f,g), it is evident that the local polarisation direction periodically swings away from $[111]_{pc}$ in Bi-enriched regions towards $[100]_{pc}$ and $[001]_{pc}$ in neighbouring Ba-enriched regions. There is also a significant reduction in the magnitude of polarisation in such regions, together with local reorientation which is driven by the interfacial and electromechanical coupling between regions with elemental segregation, analogous to similar effects in ferroelectric superlattices[28]. The alternating polarisation directions ($-[001]_{pc}-[111]_{pc}-[100]_{pc}-[111]_{pc}-$) are determined by the periodic arrangement of adjacent antiparallel electric fields induced by modulated point defect ($Bi^{\bullet}_{Ba}$ and $Ba'_{Bi}$) enrichment (i.e. valence mismatch) within the $[111]_{pc}$-orientated Ba- and Bi-enriched regions (Fig. 4b), which forms adjacent H-H and T-T domain configurations at the nanoscale along with the associated polarisation bound charge. Additional components of polarisation in the transverse (y-) direction are also evident (Fig. 4b), which counter the ionic bound charge and local electric fields induced by point defect clustering, thus minimizing the local electrostatic energies and avoiding the so-called polarisation catastrophe[29]. The development of a transverse polarisation component deviated from [111]$_{pc}$ also contributes to the formation of wavelike patterns during the element partitioning[28]. Overall, the polarisation orientation distribution is mainly centred on the [111]$_{pc}$ direction and ranges between [100]$_{pc}$ and [001]$_{pc}$ directions along with the reduction in polarisation magnitude in the plane view of Fig 4.b,g, which forms a triangular shaped distribution contour in the polar plot of the polarisation vector orientation distribution (Fig. 4h and Extended data Fig. 6). The observation of charged domain walls in the present study is driven by the elemental segregation of aliovalent $Bi^{3+}$ and $Ba^{2+}$ ions on the A-site sublattice, whereas the H-H and T-T domain configurations reported previously in $BiFeO_3$ and $PbZr_{0.2}Ti_{0.8}O_3$ systems are charge compensated by the variance of B-site oxidation state[30,31]. These observations also indicate that the signature random local electric field in relaxor ferroelectrics (as that of BFH samples shown in Fig. 2g) can be transformed to a modulated local electric field by the combination of element partitioning and domain engineering, and the charged domain walls can be created as a second order effect of aliovalent element segregation in addition to phase separation. The static ionic bound charge potential between immobile $Bi^{3+}$ and $Ba^{2+}$ enriched regions also serves to imprint the local polarisation distribution and domain configurations, which leads to a random or biased internal field at macroscopic level in BFH with and without pre-poling giving rise to anti-ferroelectric and bias hard ferroelectric characteristics respectively (Fig. 2f).

In comparison with the homogenous state of the AQ sample (Fig. 2h), the crystallographic and elemental partitioning results indicate that the formation of the nanoscale features during thermal ageing were accomplished by spinodal decomposition, in preference to the conventional nucleation and growth mechanism. This process is driven by thermodynamic immiscibility between the $BiFeO_3$ and $BaTiO_3$ components of the solid solution, leading to partial solubility and selective elemental partitioning by diffusion of A-site cations in the perovskite $ABO_3$ structure. Furthermore, the formation of dual-phase percolation networks (Fig. 4a), lattice coherency (Fig. 4b), and negligible lattice misfit (Extended data Fig. 10) between adjacent phase-separated regions indicate the development of Eshelby-type constraints[32] and opposing hydrostatic stresses within Bi- and Ba-enriched regions, considering the significant discrepancy between the ionic radii of $Ba^{2+}$(1.61 Å) and $Bi^{3+}$ (1.45 Å, extrapolated value) in 12-fold co-ordination[33]. The remarkable increment in Curie temperature for the thermally aged sample is attributed to the high Curie temperature of the Bi-enriched phase within the dual-elemental segregated superstructure, combined with tensile hydrostatic pressure exerted by the surrounding Ba-enriched regions with correspondingly larger unit cell volume. This effect is analogous to that observed in magnetic spinodal alloys[34], distinct to the biaxial strain engineered epitaxial heterostructures[4], and is opposite to that induced by compressive hydrostatic pressure or chemical pressure in $BiFeO_3$ -based systems[35].

In summary, the creation of BFH bridges the gap between ferroelectric epitaxial heterostructures and bulk polycrystalline ferroelectrics. The exploitation of interfacial valence and lattice mismatch in BFH has the potential to unleash normally inaccessible material properties in bulk ferroelectrics including: 1) transformation between associated long range ordering (normal FEs), random short range ordering (relaxor) and directional modulated local electric fields (biased hard FEs) in ferroelectrics using straightforward heat treatment without chemical modification; 2) control of polarisation continuity with locked-in bound charge; 3) formation of charged domain walls as a second order effect of aliovalent elemental partitioning; and 4) exploitation of opposing hydrostatic stress leading to enhanced Curie temperature and piezoelectric performance for emergent high temperature piezoelectric applications. Further consideration of immiscibility in ferroelectric solid solutions could also provide an alternative perspective on the evolution of ferroelectricity as a function of controlling kinetic and thermodynamic variables. The considerable scope for investigating the formation mechanisms of BFH under different electrical, mechanical and thermal boundary underlines a unique design approach to BFH, which combines domain engineering with nanoscale elemental partitioning. This approach provides a distinct and sustainable pathway to attain unprecedented functionality and stability in ferroelectrics for high temperature, high voltage, high power electronics, electro-mechanical actuators, and dielectric energy storage device applications. There are also opportunities to exploit other types of bulk heterostructures in related areas for the design of new multiferroics, ionic conductors, thermoelectrics, pyroelectrics, catalysts and many more by the exploitation of the combination of domain engineering and immiscibility in ferroic solid solutions.

**Methods**

**Sample preparation using planetary ball milling assisted solid state reaction and following heat treatment.** (1-x)BiFeO$_3$-xBaTiO$_3$ ceramics with various compositions (x=0.2, 0.25, 0.3, 0.35, 0.4) were prepared by solid state reaction, using commercially available powders of Bi$_2$O$_3$ (99.9%), Fe$_2$O$_3$ (99.9%), BaCO$_3$ (99.9%) and TiO$_2$ (99.9%) (Sigma-Aldrich). The stoichiometric mixture of the precursor powders was planetary ball milled (FRITSCH PULVERISETTE 5) with yttria-stabilized zirconia beads as milling media in propan-2-ol for 6 h using a rotational speed of 400 rpm. Afterwards, the dried powders were calcined at 800 °C for 4 h in air and wet planetary ball milled for a further 6 h. The calcined powders were dried and uniaxially pressed in a cylindrical steel die at a pressure of 50 MPa. The green pellets were then sintered at temperatures from 1010 to 1100 °C (depending upon composition) for 2 h. Planetary ball milling was previously demonstrated as an effective method to suppress the core-shell structure formation along with enhanced homogenisation during calcination and sintering. The air quenched BF-BT ceramics were prepared by reheating the as sintered samples to 900 °C (above the region of immiscibility), dwelling for 30 min and then quenching to room temperature in air. The 8 mm x 1 mm ceramic discs were transferred from 900 °C environment in furnace and placed on a alumina tile at room temperature during quenching, which allows a fast cooling for the samples to reach room temperature within 30 s, given a cooling rate about 30 °C/s. Thermal ageing was conducted on both as sintered, poled air quenched samples at different temperatures (500 – 800 °C) and with various dwell times (12 – 240 h) to create BFH(AS) and poled BHF(AQ) samples respectively. It was considered that an external short circuit was not necessary during thermal ageing of the poled sample due to the thermally activated conductivity, providing a pathway to eliminate any pyroelectric potential[36]. An additional stepwise thermal ageing experiment was carried out for poled BFH (as) samples using a poled as sintered sample by annealing at 430 °C, and subsequently at 470 °C, with a dwell time of 150 h for each step. Conductive electrodes were applied using a silver paste (GWENT C2130823D1), fired-on at a relatively low temperature of 300 °C for 10 min to minimise the thermal influence on the samples.

**Fracture toughness determination by single-edge V-notched beam (SEVNB) method.** The fracture toughness of as sintered, normal FE and tempered normal FE samples ($K_{IC}$) were determined by (SEVNB) method following the ASTM standard (ASTM C1421-18). Rectangular-ended beam specimens (12 mm × 2 mm (W) × 1.5 mm (B)) were cut from the well-sintered pellets using precision diamond wheel cutting, in order to satisfy the configuration requirements for SEVNB test. An ultra-sharp V-notch was subsequently introduced with depth along the height direction of test specimens performed with Helios 5 Laser-PFIB equipped with fully integrated femtosecond laser. At least 5 valid beams were tested for each sample with different heat treatment conditions. The fracture toughness was evaluated from the fracture load and specimen geometries as followings[37]:

$$K_{IC} = \left(\frac{P_{max} S_0 10^{-6}}{B W^{\frac{3}{2}}}\right)\left(\frac{3 \times \left(\frac{a}{W}\right)^{\frac{1}{2}} g}{2 \times \left(1 - \frac{a}{W}\right)^{\frac{3}{2}}}\right) \quad (1)$$

where g is a dimensionless correction factor given as:

$$g = g\left(\frac{a}{W}\right) = \frac{1.99 - \left(\frac{a}{W}\right)\left(1 - \frac{a}{W}\right)\left[2.15 - 3.93\left(\frac{a}{W}\right) + 2.7\left(\frac{a}{W}\right)^2\right]}{1 + 2\left(\frac{a}{W}\right)} \quad (2)$$

where P$_{max}$ is the fracture load. *a* is the depth of pre-made notch. B, W and S$_0$ refer to sample height (2 mm) and width (1.5 mm) and span width (10 mm) respectively.

**Ferroelectric, piezoelectric and dielectric property measurements.** Ferroelectric polarization-electric field (P-E) hysteresis loop measurements were performed on the BF-BT ceramics with a high voltage sinusoidal waveform at a frequency of 1 Hz, using a HP33120A function generator in conjunction with a Chevin Research HVA1B amplifier. The dielectric permittivity was measured upon heating from room temperature to 700 °C with a heating rate of 2 °C/min and at frequencies of 1, 10 and 100 kHz, using a HP 4284 A precision LCR meter. The samples were poled under a DC field of 6 kV/mm at 100 °C for 10 min. The ex-situ temperature dependence of piezoelectric coefficient $d_{33}$ was determined by quasi-static measurements of the $d_{33}$ values at room temperature with a Sinocera YE2730A $d_{33}$ meter after the samples were stabilized at multiple specified temperatures (25 - 850 °C) for 30 min followed by air quenching to room temperature. The temperature dependent piezoelectric coefficient, $d_{33}$, and coupling factor, $k_{33}$ was measured by fitting the impedance spectra with the approach proposed by Smits and Sherrit et al[38], in order to mitigate the high dielectric loss of the piezoelectrics at elevated temperatures. All the material parameters are considered as complex forms, and the loss are given by the corresponding imaginary part. The implementation of the impedance spectrum fitting was performed with in-house developed MATLAB script.

**S/TEM sample preparation and Imaging methods.** S/TEM samples were prepared using the FEI Helios Nanolab 660 focussed ion beam (FIB) scanning electron microscope (SEM) for thinning to produce a cross sectional lamella. The lamella sample was prepared using gradually decreasing ion beam milling voltages of 30 16, 8 kV for thinning, followed by, 5, 2 and 1 kV for polishing[39]. Afterwards, the TEM samples were transferred to Precision Ion Polishing System II (PIPS II) for the final polishing to eliminate the damage layer induced by FIB. Final polishing was done with an ion beam voltage of 0.3 and 0.1 kV at cryogenic temperature. This refined TEM sample preparation process plays a key role in revealing the fine spinodal decomposition patterns by removing artefacts that can be present due to FIB damage. The lamellae were prepared to be thin (less than 30 nm, confirmed by EELS). Additional needle-shape sample for TEM tomography was prepared by FIB using the same parameters mentioned above.

TEM imaging was performed on a Talos F200X (Thermo Fisher Scientific) operated at 200 kV. STEM imaging and analysis techniques were performed on the Titan G2 80-200 S/TEM ChemiSTEM (Thermo Fisher Scientific) and a double-corrected JEOL ARM300F (ePSIC, Diamond Light Source, UK) microscopes operating at 200 kV and 300 kV respectively. HAADF images were collected in the range 54-200 mrad with a probe forming a convergence semi-angle of 21 mrad. The influences of the sample thickness, sample tilting and contamination diffraction contrast was carefully minimised by the preparation of thin FIB samples (less than 30 nm with a local thickness variation less than 1 nm for the region of interests, confirmed by EELS), and accurate electron beam alignment along the zone axis to avoid parasitic angular deviation of intensity distribution within diffracted beam. For accurate atom position and displacement measurements, each atomic resolution image is reconstructed from 20 frames (2048x2048) of orthogonal fast scans, which were aligned using SmartAlign for rigid and non-rigid distortions correction. Atom column positions were determined using atom column indexing[40] and Atomap[41] with further polarisation distribution analysis being performed using an in-house Matlab script. Geometric phase analysis (GPA) was performed with Strain++[42] using a mask size roughly equal to half the reciprocal lattice parameter. The STEM-EDX spectrum imaging was acquired using the Titan's SuperX Energy Dispersive X-ray Spectroscopy (EDS) detectors and processed within the Velox software. STEM-EELS was performed using the Titan's GIF Quantum ER System with an entrance aperture of 5 mm, 0.1 s total dwell time, and a dispersion of 0.25 eV/ch.

The HAADF-STEM image series for tomographic construction was acquired with a Fischione 2020 single tilt tomography holder ranging from ±70° with a 5° intervals. Tomographic reconstruction was performed using SIRT algorithm with Etomo package. Further segmentation and visualization of the tomographic data was performed using the Thermo Scientific™ Avizo package.

fields from HREM micrographs. *Ultramicroscopy* **74**, 131–146 (1998).